\documentstyle[prl,aps,epsfig,multicol]{revtex}

\newcommand{\mc}{\multicolumn}  
\newcommand{\lsim}{\mathrel{\mathop{\kern 0pt \rlap
  {\raise.2ex\hbox{$<$}}}
  \lower.9ex\hbox{\kern-.190em $\sim$}}}
\newcommand{\gsim}{\mathrel{\mathop{\kern 0pt \rlap
  {\raise.2ex\hbox{$>$}}}
  \lower.9ex\hbox{\kern-.190em $\sim$}}}

\title{Towards the solution of
the $\Gamma_n/\Gamma_p$ Puzzle in the
Non--Mesonic Weak Decay of $\Lambda$--Hypernuclei}

\author{G. Garbarino, A. Parre\~{n}o and A. Ramos}

\address{Departament d'Estructura i Constituents de la Mat\`{e}ria,
Universitat de Barcelona, E--08028 Barcelona, Spain} \date{\today}

\begin{document}
\draft
\maketitle

\begin{abstract}

One of the main open problems in the physics of $\Lambda$--hypernuclei is the
lack of a sound theoretical interpretation of the large experimental
values for the ratio $\Gamma_n/\Gamma_p\equiv \Gamma(\Lambda n\to nn)/
\Gamma(\Lambda p\to np)$. To approach the problem, we have incorporated a 
one--meson--exchange model for the $\Lambda N\to nN$ transition in finite nuclei
in an intranuclear cascade code for the calculation 
of single and double--coincidence nucleon distributions corresponding to
the non--mesonic weak decay of $^5_\Lambda {\rm He}$ and $^{12}_\Lambda {\rm C}$. 
Due to the elimination of interferences, two--nucleon coincidences
are expected to give a cleaner determination of $\Gamma_n/\Gamma_p$
than single--nucleon observables.
Single--nucleon distributions are found to be less sensitive to variations
of $\Gamma_n/\Gamma_p$ than double--coincidence spectra. 
The comparison of our results with preliminary KEK coincidence data
allows us to extract a $\Gamma_n/\Gamma_p$ ratio for $^5_\Lambda {\rm He}$
of $0.39\pm 0.11$ when  multinucleon induced channels are omitted.

\end{abstract}

% 21.80.+a   Hypernuclei 
% 13.30.Eg   Hadronic decays
% 13.75.Ev   Hyperon-nucleon interactions 
% 25.40.-h   Nucleon-induced reactions

\begin{multicols}{2}

%%%%%%%%%%%%%%%%%%%%%%%%%%%%%%%%%%%%%%%%%%%%%%%%%%%%%%%%%%%%%%%%%%%%%%
\section{Introduction}
%%%%%%%%%%%%%%%%%%%%%%%%%
\label{intro}

An old challenge of hypernuclear decay studies has been to secure the
``elusive'' theoretical explanation of the large experimental 
values of the ratio between the neutron-- and proton--induced non--mesonic 
decay rates, $\Gamma_n/\Gamma_p\equiv \Gamma(\Lambda n\to nn)/\Gamma(\Lambda p\to np)$
\cite{Al02,Ra98}.   
Indeed, the calculations underestimate the data for
all considered hypernuclei. Moreover, in the experiments performed
until now it has not been possible to 
distinguish between nucleons produced by the one--body induced channel,
$\Lambda N\to nN$, and the two--body induced mechanism, 
$\Lambda NN\to nNN$, which is expected to be non--negligible and thus important
for the determination of $\Gamma_n/\Gamma_p$. 

Because of its strong tensor 
component, the one--pion--exchange (OPE) model with the ${\Delta}I=1/2$ isospin
rule supplies very small ratios, typically in the interval $0.05\div 0.20$. 
On the contrary, the OPE description can reproduce the total 
non--mesonic decay rates observed for light and medium hypernuclei.

Other interaction mechanisms might then be necessary to correct for the 
overestimation of $\Gamma_p$ and the underestimation of $\Gamma_n$
characteristic of the OPE.
Those which have been studied extensively in the literature are the 
following ones: i) the inclusion in the ${\Lambda}N\rightarrow nN$
transition potential of mesons heavier than the pion (also including the exchange
of correlated or uncorrelated two--pions)
\cite{Du96,Pa97,Os01,Pa01,It98}; ii) the inclusion of interaction terms that
explicitly violate the ${\Delta}I=1/2$ rule
\cite{Al02,Pa98,Al99b}; iii) the inclusion of the two--body induced decay mechanism
\cite{Al91,Ra95,Al00,Al99a} and iv) the description of the
short range $\Lambda N\to nN$ transition in terms of quark degrees of freedom
\cite{Ok99,Sa02}, which automatically introduces $\Delta I=3/2$ contributions.    

Recent progress has been made on the subject. \\
(i) On the one hand, a few calculations \cite{Os01,Pa01,It98,Ok99}
with $\Lambda N \rightarrow nN$ transition potentials including
heavy--meson--exchange and/or direct quark contributions
obtained ratios more in agreement with data, without providing, nevertheless, 
an explanation of the origin of the puzzle \cite{Al02}.
In particular, these calculations
found a reduction of the proton--induced decay width due to the 
opposite sign of the tensor component of $K$--exchange with respect to the one
for $\pi$--exchange. Moreover, the parity violating
$\Lambda N(^3S_1)\to nN(^3P_1)$ transition, which contributes to both the $n$--
and $p$--induced processes, is considerably enhanced by $K$--exchange
and direct quark mechanisms and tends to increase $\Gamma_n/\Gamma_p$ \cite{Pa01,Ok99}.
Very recently, the $\Lambda N\to nN$ interaction has been studied within an 
effective field theory framework\cite{Pa03}. The decay of $s$-- and $p$--shell
hypernuclei was approached following the same formalism as in Refs.~\cite{Pa97,Pa01},
but the weak transition was described in terms of OPE, one--kaon--exchange and 
$|\Delta S|=1$ four--fermion contact terms. The results obtained in Ref.~\cite{Pa03}
are very encouraging and open a new door for systematic studies of hypernuclear
weak decay based on effective field theory descriptions. \\
(ii) On the other hand, an error in the
computer program employed in Ref.~\cite{Ra97} to evaluate the single--nucleon 
energy spectra from non--mesonic decay has been
detected and corrected in Ref.~\cite{Ra02}. It consisted 
in the underestimation, by a factor ten, of the nucleon--nucleon collision
probabilities entering the intranuclear cascade calculation. 
The correction of such an error leads to 
quite different spectral shapes and made it possible to 
reproduce old experimental data for $^{12}_\Lambda$C \cite{Mo74,Sz91}
even with a vanishing value of $\Gamma_n/\Gamma_p$
(which is a free parameter in the polarization propagator model of
Refs.~\cite{Ra97,Ra02}). However, when compared with the recent 
proton energy spectra measured by KEK--E307 \cite{Ha01}, 
the corrected distributions still provide a quite large value of the ratio: 
$\Gamma_n/\Gamma_p=0.87\pm 0.23$ for $^{12}_\Lambda$C \cite{Sato02},
which is incompatible with pure theoretical estimations,
ranging from $0.1$ to $0.5$.

%%%%%%%%%%%%%%%%%%%%%%%%%%%
% SCOPE OF THIS PAPER
%%%%%%%%%%%%%%%%%%%%%%%%%%%   
In the light of these recent developments and of new experiments
at KEK \cite{Ou00a}, FINUDA \cite{FI01} and BNL \cite{Gi01},
it is important to develop different theoretical
approaches and strategies for the determination of $\Gamma_n/\Gamma_p$.
In a previous Letter \cite{previous}
we discussed some results of an evaluation of nucleon--nucleon
coincidence distribution in the 
non--mesonic weak decay of $^5_\Lambda {\rm He}$ and $^{12}_\Lambda$C
hypernuclei. The calculations were performed by combining a one--meson--exchange
model describing one--nucleon induced weak decays in finite nuclei 
with an intranuclear cascade code taking into account the nucleon final state 
interactions. The two--nucleon induced channel was also taken into account,
treating the nuclear finite size effects by means of a local density approximation 
scheme.

In the present paper we discuss the nucleon correlation observables
in a more systematic way. In order to stress the importance of the 
coincidence analysis in connection with the determination of $\Gamma_n/\Gamma_p$,
we also discuss single--nucleon distributions. 
In principle, the correlation observables permit a {\it cleaner}
extraction of $\Gamma_n/\Gamma_p$ from data than single--nucleon observables. 
This is due to the elimination of interference terms
between $n$-- and $p$--induced decays \cite{Al02}, which are
unavoidable in experimental data and 
cannot be taken into account by the Monte Carlo methods
usually employed to simulate the nucleon propagation through the residual nucleus. 
We also perform a weak interaction model independent analysis to extract an estimate
for $\Gamma_n/\Gamma_p$ using preliminary results from KEK \cite{Ou00a,OuOk}
on two--nucleon angular and energy correlations.
The resulting $\Gamma_n/\Gamma_p$ values for $^5_\Lambda {\rm He}$
turn out to be substantially smaller than those obtained in the past
\cite{Sz91,No95a} from single coincidence analyses and fall within the predictions 
of recent theoretical studies \cite{Pa01,It98,Ok99}. 

The work is organized as follows. In Sec.~\ref{models} we give an
outline of the models employed to describe the non--mesonic weak decay. 
In Sec.~\ref{mc} we discuss the main features of the
Monte Carlo simulation accounting for the nucleon propagation inside 
the residual nucleus. The purpose of Sec.~\ref{coinc} is to discuss 
the reasons why multi--nucleon coincidence analyses should be preferred
over single--nucleon studies in order to extract $\Gamma_n/\Gamma_p$ from data.
Our results are discussed in Sec.~\ref{res} and the conclusions are given
in Sec.~\ref{conc}. 

%*********************************************************************
\section{Models for the weak decay} 
%*********************************************************************  
\label{models}
\subsection{One--nucleon induced decay: finite nucleus approach}
%*********************************************************************
\label{fn}

The one--nucleon induced non--mesonic decay rate can be written as:
\begin{eqnarray}
\label{gamma1}
\Gamma_1 &=& \int  \frac{d^3 P_T}{(2\pi)^3}\int \frac{d^3 k_r}{(2\pi)^3}\,
(2\pi)\, \delta(M_H-E_R-E_1-E_2) \nonumber \\ 
&&\times \frac{1}{(2J+1)} \sum_{^{M_J  \{R\} }_{\{1\} \{2\}}}
\mid {\cal M}_{fi} \mid^2  , \nonumber
\end{eqnarray}  
where the initial hypernucleus, of mass $M_H$, is assumed to be at rest
and the quantities $E_R$ and $E_{1,2}$ are 
the total energy of the residual nuclear system and the total
asymptotic energies of the emitted nucleons, respectively.
The integration variables ${\vec P}_T\equiv {\vec k}_1+{\vec k}_2$ 
and ${\vec k}_r\equiv ({\vec k}_1-{\vec k}_2)/2$ are the total and relative 
momenta of the two outgoing nucleons.
The momentum conserving delta function has been used to
integrate out the momentum of the residual nucleus, 
${\vec k}_R=-{\vec P}_T$.
The sum, together with the factor $1/{(2J+1)}$, indicates an average
over the initial hypernucleus total spin
projections, $M_J$, and a sum over all quantum numbers
of the residual system, $\{R\}$, as well as over the spin and isospin
projections of the outgoing nucleons, $\{1\}$ and $\{2\}$.
Finally:
\[
{\cal M}_{fi} =
\langle \Psi_R ; {\vec P}_T \, {\vec k}_r, S\, M_S, T\, T_3| 
\hat{O}_{\Lambda N \to nN} \left| \Psi_H \right\rangle \nonumber
\]
is the amplitude for the transition from an initial hypernuclear state 
$\Psi_H$ into a final state which is factorized into 
an antisymmetrized two--nucleon state
and a residual nuclear state $\Psi_R$.
The two--nucleon state is characterized
by the total momentum ${\vec P}_T$, the relative momentum ${\vec k}_r$,
the spin and spin projection $S, M_S$ and the isospin and isospin projection
$T, T_3$. Finally, $\hat{O}_{\Lambda N \to nN}$ is a two--body transition operator
acting on all possible $\Lambda N$ pairs.  

Working in a coupled two--body spin and isospin basis,
the non--mesonic decay rate can be written as:
\[
\Gamma_1=\Gamma_n+\Gamma_p , \nonumber
\]
where $\Gamma_n$ and $\Gamma_p$ stand for the neutron--
($\Lambda n \to nn$) and proton--induced ($\Lambda p
\to np$) decay rate, respectively. They are given by ($N=n,p$):
\begin{eqnarray}
\label{gamma3}
&&\Gamma_{N} = \int \frac{d^3 P_T}{(2\pi)^3} \, \int \frac{d^3k_r}{(2\pi)^3}\,
(2\pi) \, \delta(M_H-E_R-E_1-E_2) \nonumber \\
&&\times \sum_{S M_S} \sum_{J_R M_R} \sum_{T_R T_{3_R}}
\frac{1}{2J+1} \sum_{M_J} \mid \langle T_R T_{3_R}, \frac{1}{2} t_{3N} 
\mid T_I T_{3_I} \rangle \mid^2  \nonumber \\
&&\times \biggl| \, \sum_{T T_3}
\langle T T_3 \mid \frac{1}{2} -\frac{1}{2}, \frac{1}{2} t_{3N} \rangle
\sum_{m_\Lambda M_C} \langle j_\Lambda m_\Lambda, J_C M_C \mid
J M_J \rangle \nonumber \\
&&\times \sum_{j_N} \sqrt{S \, ( J_C \, T_I \, ; J_R \, T_R \, , j_N \,t_{3N} )} \\
&&\times \sum_{M_R m_N} \langle J_R M_R, j_N m_N \mid J_C M_C\rangle \nonumber \\
&&\times \sum_{m_{l_N} m_{s_N}} \langle j_N m_N \mid l_N m_{l_N},
\frac{1}{2} m_{s_N} \rangle  \nonumber \\
&&\times \sum_{m_{l_\Lambda} m_{s_\Lambda}} \,\,
\langle j_\Lambda m_\Lambda \mid l_\Lambda m_{l_\Lambda}, \frac{1}{2}
m_{s_\Lambda} \rangle \nonumber \\
&&\times \sum_{S_0 M_{S_0}} \langle S_0 M_{S_0} \mid \frac{1}{2} m_{s_\Lambda}, \frac{1}{2}
m_{s_N} \rangle \nonumber \\
&&\times \sum_{T_0 T_{3_0}} \langle T_0 T_{3_0} \mid \frac{1}{2}\, -\frac{1}{2},
\frac{1}{2} t_{3N} \rangle \, \frac{1-(-1)^{(L+S+T)}}{\sqrt{2}} 
\nonumber \\
&&\times  t_{\Lambda N \to nN}
(S,M_S,T,T_3,S_0,M_{S_0},T_0,T_{3_0},l_\Lambda,l_N,{\vec P}_T, 
{\vec k}_r) \biggl|^2\nonumber
\end{eqnarray}
where $t_{3p}=1/2$ and $t_{3n}=-1/2$. The elementary amplitude
$t_{\Lambda N \to nN}$ accounts for the transition
from an initial $\Lambda N$ state with spin (isospin) $S_0$ $(T_0)$
to a final antisymmetric $nN$ state with spin (isospin) $S$ $(T)$. It
can be written in terms of other elementary amplitudes which depend on center--of--mass
($``R"$) and relative ($``r"$) orbital angular momentum quantum numbers
of the $\Lambda N$ and $nN$ systems:
\begin{eqnarray}
\label{moshinski}
&&t_{\Lambda N \to nN} \\
&&=\sum_{N_r L_r N_R L_R} X(N_r L_r, N_R L_R, l_\Lambda l_N)
\, t_{\Lambda N \to nN}^{N_r L_r\, N_R L_R} , \nonumber
\end{eqnarray}
where the dependence on the spin and isospin quantum numbers
has to be understood. In Eq.~(\ref{moshinski}), the coefficients 
$X(N_r L_r, N_R L_R, l_\Lambda l_N)$ are the well known Moshinsky brackets, while:
\begin{eqnarray}
\label{erre-dep}
&&t_{\Lambda N \to nN}^{N_r L_r\, N_R L_R} =
\frac{1}{\sqrt{2}}\int d^3R \int d^3r \, {\rm e}^{- i {\vec P}_T\cdot {\vec R}}
\Psi^*_{{\vec k}_r}\, ({\vec r}\,) \chi^{\dagger\, S}_{M_S}
\chi^{\dagger\, T}_{T_3} \nonumber \\
&&\times V_{\sigma,  \tau}({\vec r}\,) \, \Phi^{\rm CM}_{N_R L_R}\left(\frac{{\vec
R}}{b/\sqrt{2}}\right) \Phi^{\rm rel}_{N_r L_r}\left(\frac{{\vec r}}{\sqrt{2}b}\right)
\chi^{S_0}_{M_{S_0}} \chi^{T_0}_{T_{3_0}} .
\end{eqnarray}  
Here, $V_{\sigma, \tau}({\vec r}\,)$ stands for the one--meson--exchange weak potential,
which depends on the relative distance between the interacting $\Lambda$
and nucleon as well as on their spin and isospin quantum numbers.
Moreover, $\Phi^{\rm rel}_{N_r L_r}({\vec r}/(\sqrt{2}b))$ and
$\Phi^{\rm CM}_{N_R L_R} ({\vec R}/(b/\sqrt{2}))$ are the
relative and center--of--mass harmonic oscillator wave functions 
describing the $\Lambda N$ system, while
$\Psi_{{\vec k}_r}\, ({\vec r}\,)$ is the relative wave function
of the $nN$ final state. Further details regarding notation can be found in Ref.~\cite{Pa97}.

In order to study nucleon--nucleon coincidence spectra, it is convenient to 
introduce differential decay widths depending on the center--of--mass coordinate 
$\vec R$, the cosinus of the angle between nucleon 1 and 2, ${\rm cos} \, \theta_{12}$,
and the modulus of the momentum of one of the outgoing nucleons, say $k_1$. 
For this purpose, taking into account Eqs.~(\ref{moshinski}) and (\ref{erre-dep}),
let us rewrite the decay rate of Eq.~(\ref{gamma3}) in the following schematic way:
\begin{eqnarray}
\label{differ}
\Gamma_{N} &=& \int d^3 P_T \int d^3k_r\, \sum_i a_i \\
&&\times \left[\sum_j c_j \int d^3R \, A_{ij}(\vec{R},\vec{P}_T,\vec{k}_r) \right] \nonumber \\
&&\times \left[ \sum_{j^\prime} c^*_{j^\prime} \int d^3R^\prime 
A_{ij^\prime}^*(\vec{R}^\prime,\vec{P}_T,\vec{k}_r) \right]  \nonumber \\
%&=& \int \frac{d^3 P_T}{(2\pi)^3} \int \frac{d^3k_r}{(2\pi)^3}
%\int d^3R \, \nonumber \sum_i a_i \\
%&&\times \left[\sum_j c_j A_{ij}(\vec{R},\vec{P}_T,\vec{k}_r) \right] \nonumber \\
%&& \times \left[ \sum_{j^\prime} c^*_{j^\prime} 
%\int d^3R^\prime A_{ij^\prime}^*(\vec{R}^\prime,\vec{P}_T,\vec{k}_r) \right] \nonumber \\
&\equiv & \int d^3 P_T \int d^3k_r \int d^3R \,\, 
\Gamma_N ({\vec R}, \vec{P}_T, \vec{k}_r) , \nonumber
\end{eqnarray}
where $a_i$ and $c_j$ include Clebsh-Gordan coefficients and other factors that
depend on various quantum numbers, while $A_{ij}(\vec{R},\vec{P}_T,\vec{k}_r)$
denote $\Lambda N\to nN$ weak decay amplitudes. Changing variables from 
$\vec{P}_T,\vec{k}_r$ to $\vec{k}_1,\vec{k}_2$, using the
energy conserving delta function and imposing rotational invariance, it is possible to
substitute, in the integrand of Eq.~(\ref{differ}), the $\vec{k}_r$ and $\vec{P}_T$
dependence by a dependence on $k_1$ and ${\rm cos} \, \theta_{12}$.  
Without performing the $\vec R$, $k_1$,
and ${\rm cos} \, \theta_{12}$ integrations, one can then write
$n$-- and $p$--stimulated differential decay rates:
\begin{eqnarray}
\label{differ1}
&&\Gamma_N ({\vec R}, k_1, {\rm cos} \, \theta_{12})=
\sum_i a_i \left[\sum_j c_j A_{ij}(\vec{R},k_1,{\rm cos} \, \theta_{12}) \right] \nonumber \\
&&\hskip 1.5cm \times \left[ \sum_{j^\prime} c^*_{j^\prime}
\int d^3R^\prime A_{ij^\prime}^*(\vec{R}^\prime,k_1,{\rm cos} \, \theta_{12}) \right] ,
\end{eqnarray}
appropriate for obtaining the distributions of the weak decay nucleons
required as input of the intranuclear cascade calculation.  
As explained in the following Section, the intranuclear cascade code allows 
then the primary nucleons to change energy, direction and charge, exciting 
other secondary nucleons 
that are also followed as they move through the nucleus.

Note that by considering the primary nucleons as emerging from the same point 
$\vec R$ in space,
we have implicitly assumed that the weak transition proceeds as a point interaction.
This is implied by the shape of the regularized potentials of the
model of Refs.~\cite{Pa97,Pa01} ---shown in Ref.~\cite{Go97}---,
which peak at relative distances $r\equiv |{\vec r}_1-{\vec r}_2|\lsim 0.5$ fm.

Finally, we recall that in our calculations
the one--meson--exchange (OME) weak transition potential
entering Eq.~(\ref{erre-dep}) 
contains the exchange of $\rho$, $K$, $K^*$, $\omega$ and 
$\eta$ mesons in addition to the pion \cite{Pa01}. 
The strong couplings and strong final state interactions acting
between the weak decay nucleons are taken into account by using a scattering $nN$ 
wave function from the Lippmann--Schwinger ($T$--matrix) equation obtained with 
NSC97 (versions ``a'' and ``f'') potentials \cite{nsc}. 
The corresponding decay rates for $^5_\Lambda {\rm He}$ and $^{12}_\Lambda {\rm C}$
are listed in Table~\ref{gammas} (OMEa and OMEf)
together with the one--pion--exchange predictions (OPE).

%*******************************************************************
\subsection{Two--nucleon induced decay: local density approximation}
%*******************************************************************
\label{2b}

The differential and total decay rates for the two--nucleon induced process, 
$\Lambda np\to nnp$, are calculated with the 
polarization propagator method in local density approximation
(LDA) of Refs.~\cite{Ra95,Al00}. 
In such a calculation, the simple OPE mechanism,
supplemented by strong $\Lambda N$ and $nN$ short range correlations (given 
in terms of phenomenological Landau functions) described the weak 
transition process. The two--nucleon induced differential decay width
$\Gamma_2 ({\vec R}, k_1, {\rm cos} \, \theta_{12})$ is 
obtained from the two--particle two--hole (2p2h)
contributions to the pion self--energy. As explained in 
Ref.~\cite{Ra95}, these contributions are derived in a phenomenological way from 
a fit to pionic atoms, conveniently modified by the Lorentz--Lorenz correction 
and extended to all kinematical regions using the appropriate phase--space.
The intranuclear cascade code then follows the fate of the two nucleons excited 
by these 2p2h components, as well as that of the third nucleon emitted at the 
$\Lambda N \pi$ vertex, as they move through the nucleus. 

In the present calculation, the distributions of the weak decay nucleons and the 
value of $\Gamma_2$ have been properly scaled to maintain the ratio
$\Gamma_2/\Gamma_1$ unchanged: we then use 
$\Gamma_2/\Gamma_1=(\Gamma_2/\Gamma_1)^{\rm LDA}=0.20$ for $^5_\Lambda$He and
$0.25$ for $^{12}_\Lambda$C.
These values of $(\Gamma_2/\Gamma_1)^{\rm LDA}$
have been obtained with the latest update of parameters given in 
Ref.~\cite{Al00} after
correcting a small (conceptual) error in the implementation of data
on the $P$--wave pion--nucleus optical potential. For the hypernuclei
treated in the present paper,
$^5_\Lambda {\rm He}$ and $^{12}_\Lambda {\rm C}$, such a correction
slightly decreases the values of $\Gamma_1$ while increasing $\Gamma_2$
by about 20\%.

%****************************************
\section{Intranuclear cascade simulation}
%****************************************  
\label{mc}

In their way out of the nucleus, the weak decay (i.e., primary) nucleons
continuously change energy, direction and charge due to collisions with other nucleons. 
As a consequence, secondary nucleons are also emitted. 

We simulate the nucleon propagation inside the
residual nucleus with the Monte Carlo code of Refs.~\cite{Ra97,Ra02}.
A random number generator determines the decay channel, $n$--, $p$--
or two--nucleon induced, according to the ratios $\Gamma_n/\Gamma_p$
and $\Gamma_2/\Gamma_1$ predicted by our finite nucleus and LDA
approaches. Positions, momenta 
and charges of the weak decay nucleons are selected by the same
random number generator, according to the corresponding
probability distributions given by the finite nucleus and LDA 
calculations. 

For neutron--, proton-- and two--nucleon induced
decays, the discussion of Section~\ref{models} allows us to introduce
the differential decay rates $\Gamma_n ({\vec R}, k_1, {\rm cos} \, \theta_{12})$,
$\Gamma_p ({\vec R}, k_1, {\rm cos} \, \theta_{12})$ and 
$\Gamma_2 ({\vec R}, k_1, {\rm cos} \, \theta_{12})$
supplying the $n$--, $p$-- and two--nucleon stimulated total rates $\Gamma_n$, $\Gamma_p$ 
and $\Gamma_2$ through the following relation:
\[
\Gamma_i=\int d^3 k_1 \int d\, {\rm cos}\, \theta_{12} \int d^3 R \,\,
\Gamma_i ({\vec R}, k_1, {\rm cos} \, \theta_{12}) . \nonumber
\]

After they are produced as explained,
the primary nucleons move under a local potential
$V_N(R)=-k_{F_N}^2(R)/2 m_N$, where 
$k_{F_N}(R)=[3\pi^2\rho_N(R)]^{1/3}$ ($N=n,p$) is the
local nucleon Fermi momentum corresponding to the nucleon density $\rho_N(R)$. 
The primary nucleons also collide
with other nucleons of the medium according to free space
nucleon--nucleon cross sections \cite{Cu96} properly corrected 
to take into account the Pauli blocking effect. For further
details concerning the intranuclear cascade calculation see 
Ref.~\cite{Ra97}. Each Monte Carlo event will then end
with a certain number of nucleons which leave the nucleus
along defined directions and with defined energies. One can then 
select the outgoing nucleons and store them in different ways,
as we shall do in the discussion of Section~\ref{res}.

We are aware of the fact that accounting for nucleon final state
interactions effects in light residual nuclei 
(as those required to treat $^5_\Lambda$He) through
Monte Carlo techniques is questionable. However, realistic calculations for few--body
scattering states have been performed up to 3 nucleons only (see for
instance Ref.~\cite{Kam}).
For the hypernuclear non--mesonic decay problem,
only the case of $^3_\Lambda$H has been treated exactly \cite{Go97,Gol}.
Although one might attempt three--cluster type calculations, this goes beyond the
scope of the present work. For this reason, the results 
we present for $^5_\Lambda$He should be considered less realistic than the
corresponding ones for $^{12}_\Lambda$C. 

%***************************************************
\section{Single--nucleon vs nucleon--coincidence analyses}
%***************************************************
\label{coinc}

In this Section we discuss the reason 
why multi--nucleon coincidence studies
should be preferred over single--nucleon analyses when the purpose
is the determination of $\Gamma_n/\Gamma_p$ .
The argument for this explanation is
based on the elimination of quantum mechanical {\it interferences}
between $n$-- and $p$--stimulated weak decays \cite{Al02}. 

Let us first note that the nucleons originating from $n$-- and 
$p$--induced processes are added {\it incoherently} (i.e., classically) in 
our intranuclear cascade calculation (for the moment we are then neglecting
an analogous effect due to the two--nucleon stimulated decay channel).
However, for particular kinematics of the detected
nucleons (for instance at low kinetic energies), an in principle
possible quantum--mechanical interference effect between $n$-- and
$p$--induced channels should inevitably affect the observed distributions. 
Therefore, extracting the ratio $\Gamma_n/\Gamma_p$ from experimental data 
with the help of a classical intranuclear cascade 
calculation may not be a clean task.

To clarify better the issue, 
let us consider for instance an experiment (such as the majority
of the experiments performed up to now) measuring
{\it single--proton} kinetic energy spectra. The relevant quantity
is then the number of outgoing protons observed as a function of the 
kinetic energy $T_p$. Schematically, this observable can be written as:
\begin{eqnarray}
\label{interf}   
&&N_p(T_p)\propto
\left|\langle p(T_p)|\hat{O}_{\rm FSI}\, \hat{O}_{\rm WD}|\Psi_H \rangle \right|^2 \\
&&= \left|\alpha\, \langle p(T_p)|\hat{O}_{\rm FSI}|nn, \Psi_R\rangle
+\beta\, \langle p(T_p)|\hat{O}_{\rm FSI}|np, \Psi_{R^\prime} \rangle \right|^2 , \nonumber
\end{eqnarray} 
where $|p(T_p)\rangle$ represents a many--nucleon final state with a proton
whose kinetic energy is $T_p$. Moreover, in Eq.~(\ref{interf}) the action of 
the weak decay operator $\hat{O}_{\rm WD}\equiv \hat{O}_{\Lambda n\to nn}
+\hat{O}_{\Lambda p\to np}$ produced
the superposition:
\[
\hat{O}_{\rm WD}|\Psi_H\rangle = \alpha \, |nn, \Psi_R\rangle +
\beta \, |np, \Psi_{R^\prime} \rangle . \nonumber
\]
Here $|nn, \Psi_R\rangle$ ($|np, \Psi_{R^\prime} \rangle$) is a state with
a $nn$ ($np$) primary pair moving inside a residual nucleus $\Psi_R$ ($\Psi_{R^\prime}$). 
Note that in the present schematic picture: $\Gamma_n\propto |\alpha|^2$ and
$\Gamma_p\propto |\beta|^2$.
Since both transition amplitudes entering the last equality of
Eq.~(\ref{interf}) are in general 
non--vanishing, interference terms between $n$-- and
$p$--induced decays are expected to contribute to $N_p(T_p)$.
An amplitude $\langle p(T_p)|\hat{O}_{\rm FSI}|nn, \Psi_R\rangle$
different from zero means that, due to nucleon final state interactions (FSI), 
a secondary proton has a non--vanishing probability 
to emerge from the nucleus with kinetic energy $T_p$ even if the 
weak process was $n$--induced (i.e., without primary protons).
While for high kinetic energies (say for $T_p>80$ MeV) this amplitude is 
expected to be almost vanishing, as long as $T_p$ decreases
its contribution could produce an important interference effect 
(see the results discussed in Section~\ref{res-sing}).

An interference--free observation would imply the measurement of all
the quantum numbers of the final nucleons and residual nucleus. While this is an
impossible experiment, what is certain is that the magnitude of the interference
can be reduced if one measures in a more accurate way the final state. 
For this reason, two--nucleon coincidence observables are expected to be 
less affected by interferences than single--nucleon ones and thus more reliable for 
determining $\Gamma_n/\Gamma_p$. If we consider the detection 
of the kinetic energy {\it correlation}
of a $np$ pair emitted nearly back--to--back (say with ${\rm cos}\, \theta_{np}\leq -0.9$):
\begin{eqnarray}
\label{interf2}
&&N_{np}(T_n+T_p,{\rm cos}\, \theta_{np}\leq -0.9) \\
&&\propto \left|\alpha \langle n(T_n), p(T_p), {\rm cos}\, \theta_{np}\leq -0.9|\hat{O}_{\rm FSI}
|nn, \Psi_R\rangle \right. \nonumber \\
&&\left. +\beta \langle n(T_n), p(T_p), {\rm cos}\, \theta_{np}\leq -0.9
|\hat{O}_{\rm FSI}|np, \Psi_{R^\prime} \rangle \right|^2 , \nonumber
\end{eqnarray}
one expects an interference effect smaller than the one appearing in
single--nucleon observations, i.e., in Eq.~(\ref{interf}),
especially when particular energy cuts are considered.

To see this more explicitly, let us rewrite Eqs.~(\ref{interf}) and (\ref{interf2}) in the
following way:
\begin{eqnarray}
\label{single}
N_p&=&\left[N^{\Lambda n\to nn}_p+N^{\Lambda p\to np}_p\right]\left[1+I_p\, {\rm cos}\, \phi_p\right] ,\\
\label{double}
N_{np}&=&\left[N^{\Lambda n\to nn}_{np}+N^{\Lambda p\to np}_{np}\right]\left[1+I_{np}\, {\rm cos}\, \phi_{np}\right] ,
\end{eqnarray}
where the various arguments have been suppressed and the number of nucleons:
\begin{eqnarray}
N^{\Lambda n\to nn}_p&\equiv& \left|A^{\Lambda n\to nn}_p\right|^2
\propto \left|\alpha\right|^2\left|\langle p|\hat{O}_{\rm FSI}|nn,\Psi_R\rangle\right|^2 , \nonumber \\
N^{\Lambda p\to np}_p&\equiv& \left|A^{\Lambda p\to np}_p\right|^2 
\propto \left|\beta\right|^2\left|\langle p|\hat{O}_{\rm FSI}|np,\Psi_{R'}\rangle\right|^2 , \nonumber \\
N^{\Lambda n\to nn}_{np}&\equiv& \left|A^{\Lambda n\to nn}_{np}\right|^2 
\propto \left|\alpha\right|^2\left|\langle np|\hat{O}_{\rm FSI}|nn,\Psi_R\rangle\right|^2 , \nonumber \\
N^{\Lambda p\to np}_{np}&\equiv& \left|A^{\Lambda p\to np}_{np}\right|^2 
\propto \left|\beta\right|^2\left|\langle np|\hat{O}_{\rm FSI}|np,\Psi_{R'}\rangle\right|^2 , \nonumber 
\end{eqnarray}
have obvious meaning. Moreover:
\begin{eqnarray}
\label{single-vis}
I_p&=& \frac{2 \sqrt{N^{\Lambda n\to nn}_p\, N^{\Lambda p\to np}_p}}
{N^{\Lambda n\to nn}_p+N^{\Lambda p\to np}_p} , \nonumber \\
\label{double-vis}
I_{np}&=& \frac{2 \sqrt{N^{\Lambda n\to nn}_{np}\, N^{\Lambda p\to np}_{np}}}
{N^{\Lambda n\to nn}_{np}+N^{\Lambda p\to np}_{np}} , \nonumber
\end{eqnarray}
and $\phi_p=\phi^{\Lambda n\to nn}_p-\phi^{\Lambda p\to np}_p$ 
($\phi_{np}=\phi^{\Lambda n\to nn}_{np}-\phi^{\Lambda p\to np}_{np}$) is the phase difference
between the amplitudes $A^{\Lambda n\to nn}_p=\sqrt{N^{\Lambda n\to nn}_p}\, e^{i\phi^{\Lambda n\to nn}_p}$ 
and $A^{\Lambda p\to np}_p=\sqrt{N^{\Lambda p\to np}_p}\, e^{i\phi^{\Lambda p\to np}_p}$ 
($A^{\Lambda n\to nn}_{np}=\sqrt{N^{\Lambda n\to nn}_{np}}\, e^{i\phi^{\Lambda n\to nn}_{np}}$
and $A^{\Lambda p\to np}_{np}=\sqrt{N^{\Lambda p\to np}_{np}}\, e^{i\phi^{\Lambda p\to np}_{np}}$).
Note that the distributions of Eq.~(\ref{single}) [Eq.~(\ref{double})] are not affected by interference
only when $I_p\,{\rm cos}\,\phi_p=0$ [$I_{np}\,{\rm cos}\,\phi_{np}=0$].
Since, as explained, we expect 
$N^{\Lambda n\to nn}_{np}/N^{\Lambda p\to np}_{np}< 
N^{\Lambda n\to nn}_p/N^{\Lambda p\to np}_p<1$, then $I_{np}$ will be smaller than $I_p$.
The numerical results
discussed in Sections~\ref{res-sing} and \ref{res-coin} confirm this expectation.

The same reasoning must be applied ---and the previous equations changed accordingly---
once the two--nucleon induced decay mechanism is taken into account.
Note, however, that our discussion in this Section have
neglected the effect of the phase differences $\phi_p$ and $\phi_{np}$,
which cannot be evaluated theoretically.  
A quite indirect estimate of these phases (and then of the real interference effects)
can be obtained through the comparison
of our calculated spectra (in which interference is not taken into account)
with the experimental ones (which can be affected by interferences).

%%%%%%%%%%%%%%%%%
\section{Results}
%%%%%%%%%%%%%%%%%
\label{res}

%************************************
\subsection{Single--nucleon spectra}
%***********************************
\label{res-sing}

\begin{figure}
\begin{center}
\mbox{\epsfig{file=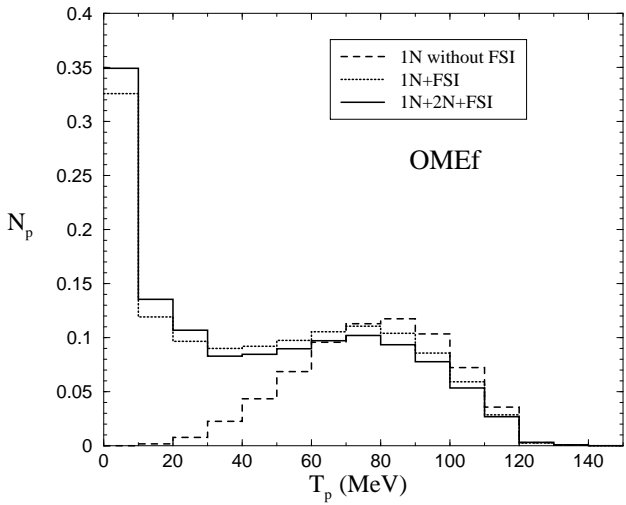,width=.45\textwidth}}
%\mbox{\epsfig{file=s1.eps}}
\vskip 2mm
\caption{Single--{\it proton} kinetic energy spectra for the non--mesonic
decay of $^5_\Lambda$He. The dashed and dotted lines are normalized {\it per
one--nucleon induced decay} ($\Gamma_1=\Gamma_n+\Gamma_p$),
while the continuous line is normalized
{\it per non--mesonic decay} ($\Gamma_{\rm NM}=\Gamma_1+\Gamma_2)$.}
\label{s1}
\end{center}
\end{figure}

\begin{figure}
\begin{center}
\mbox{\epsfig{file=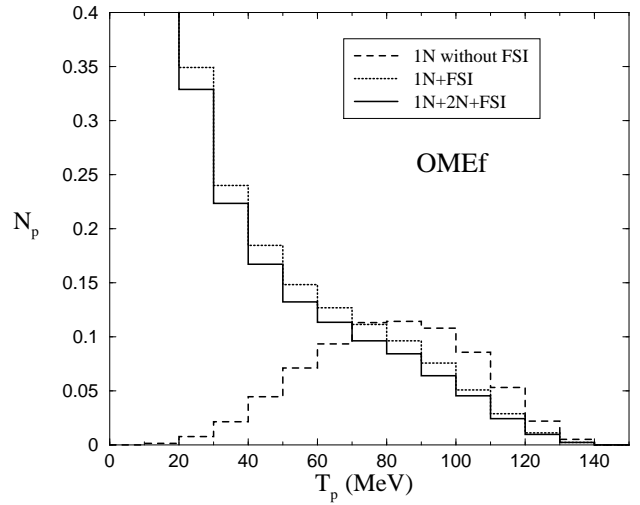,width=.45\textwidth}}
%\mbox{\epsfig{file=s2.eps}}
\vskip 2mm
\caption{Same as in Fig.~\ref{s1} for $^{12}_\Lambda$C.}
%\caption{Single--{\it proton} kinetic energy spectra {\it per non--mesonic
%weak decay} of $^{12}_\Lambda$C.}
\label{s2}
\end{center}
\end{figure}

In Figs.~\ref{s1} and \ref{s2} we show the proton kinetic energy spectra corresponding
to the decay of $^5_\Lambda$He and $^{12}_\Lambda$C, respectively. 
Note the particular normalizations of the curves presented in this and in the
following figures. The dashed curves correspond to the distributions of the 
one--nucleon induced primary protons:
they have been obtained through intranuclear cascade calculations in which the 
one--nucleon stimulated weak decay nucleons move under the effect of the 
nuclear local potential, $V_N(R)=- k^2_{F_N}(R)/2 m_N$, without colliding
with the nucleons of the medium. The inclusion of nucleon FSI provide the
result given by the dotted lines. The continuous lines correspond to the full
calculation, i.e., once the two--nucleon induced channel
is also included. The calculations
have been performed with the OMEf model, which predicts $\Gamma_n/\Gamma_p=0.46$
for $^5_\Lambda$He and $\Gamma_n/\Gamma_p=0.34$ for $^{12}_\Lambda$C. 
The model OMEa supplies similar results both for the proton and neutron spectra.
While the primary proton distributions 
are very similar for the two hypernuclei (the one for
$^{12}_\Lambda$C is slightly broader), the full calculations clearly
show that, as expected, FSI have a bigger effect for the heavier
system. For $^{12}_\Lambda$C, FSI are so important that
they completely smear out the maxima corresponding to the primary 
protons. Our predictions for the proton spectra from $^{12}_\Lambda$C
should be compared 
with the one measured at KEK--E307 \cite{Ha01}. Unfortunately, this is not
possible since these data
have not been corrected for the detector geometry and the
energy losses occurring in the target and detector materials.  


Let us now introduce the number of nucleons of the type $N$ ($N=n,p$) produced in $n$--induced
($N^{\rm 1Bn}_N$), $p$--induced ($N^{\rm 1Bp}_N$) and two--nucleon induced 
($N^{\rm 2B}_N$) decays. If we normalize these quantities per
$n$--induced, $p$--induced, and two--nucleon--induced decay, respectively,
the {\it total} number of nucleons of the type $N$ is given by:
\begin{eqnarray}
\label{single-n}
N_N&=&\frac{N^{\rm 1Bn}_N\, \Gamma_n+N^{\rm 1Bp}_N\, 
\Gamma_p+N^{\rm 2B}_N\, \Gamma_2}
{\Gamma_n+\Gamma_p+\Gamma_2} \\
&\equiv&N^{\Lambda n\to nn}_N+N^{\Lambda p\to np}_N 
+N^{\Lambda np\to nnp}_N, \nonumber
\end{eqnarray}  
where $N^{\Lambda n\to nn}_N$, $N^{\Lambda p\to np}_N$ and 
$N^{\Lambda np\to nnp}_N$ have obvious meaning.
All these nucleon numbers can be considered either as being functions of the nucleon 
kinetic energy, $N_N(T_N)$, as it is the case of Figs.~\ref{s1} and \ref{s2},
or as the corresponding integrated quantities,
$N_N=\int dT_N N_N(T_N)$. 
By construction, $N^{\rm 1Bn}_N$, $N^{\rm 1Bp}_N$ and $N^{\rm 2B}_N$
($N^{\Lambda n\to nn}_N$, $N^{\Lambda p\to np}_N$ and $N^{\Lambda np\to nnp}_N$)
{\it do not} depend ({\it do} depend) on the interaction model 
employed to describe the weak decay.

\begin{figure}
\begin{center}
\mbox{\epsfig{file=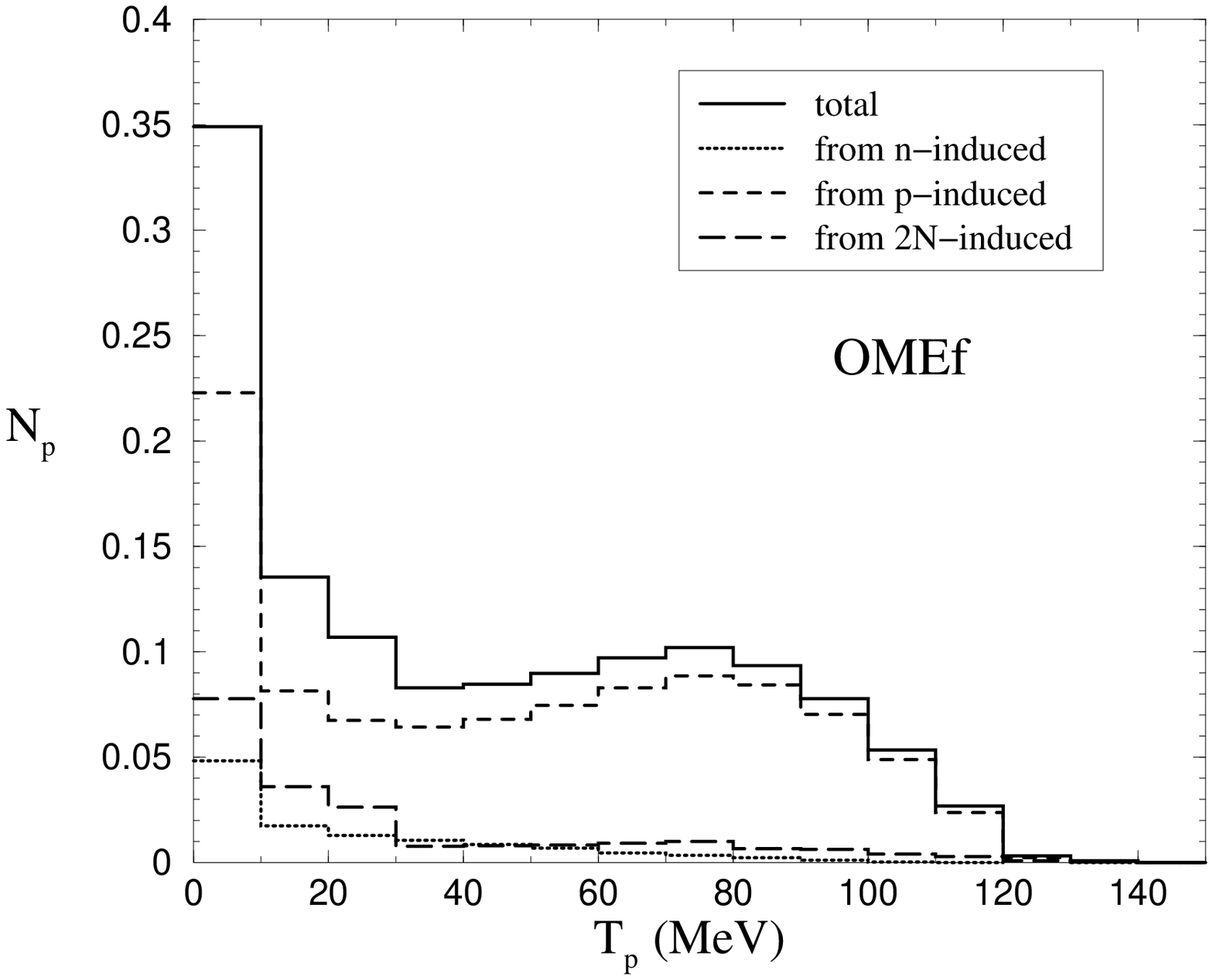,width=.45\textwidth}}
%\mbox{\epsfig{file=s3p.eps}}
\vskip 2mm
\caption{Single--{\it proton} kinetic energy spectra for the non--mesonic
weak decay of $^5_\Lambda$He. The total spectrum $N_p$ (normalized
{\it per non--mesonic weak decay}) has been decomposed
in its components $N^{\Lambda n\to nn}_p$,
$N^{\Lambda p\to np}_p$ and $N^{\Lambda np\to nnp}_p$ according to
Eq.~(\ref{single-n}).}
\label{s3p}
\end{center}
\end{figure}

\begin{figure}
\begin{center}
\mbox{\epsfig{file=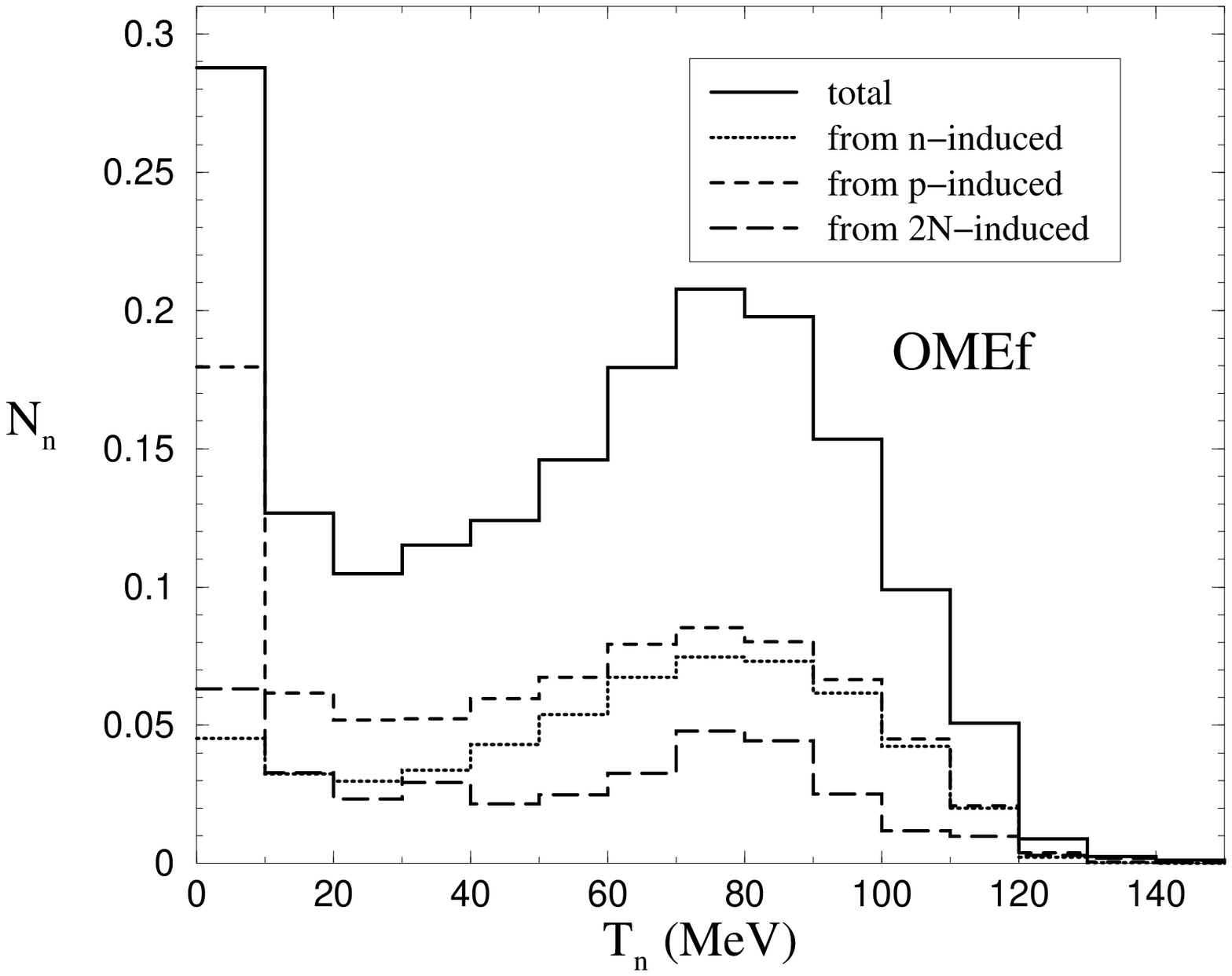,width=.45\textwidth}}
%\mbox{\epsfig{file=s3n.eps}}
\vskip 2mm
\caption{Single--{\it neutron} kinetic energy spectra for the non--mesonic
weak decay of $^5_\Lambda$He. The total spectrum $N_n$ (normalized
{\it per non--mesonic weak decay}) has been decomposed
in its components $N^{\Lambda n\to nn}_n$,
$N^{\Lambda p\to np}_n$ and $N^{\Lambda np\to nnp}_n$ according to
Eq.~(\ref{single-n}).}
\label{s3n}
\end{center}
\end{figure}

\begin{figure}
\begin{center}
\mbox{\epsfig{file=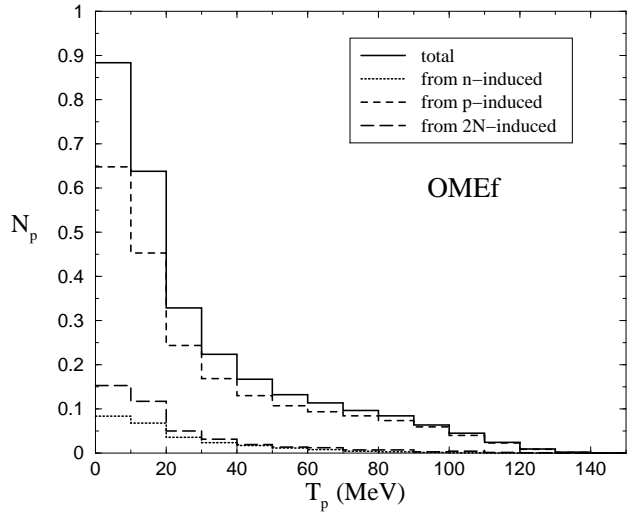,width=.45\textwidth}}
%\mbox{\epsfig{file=s4p.eps}}
\vskip 2mm
\caption{Same as in Fig.~\ref{s3p} for $^{12}_\Lambda$C.}
%\caption{Single--{\it proton} kinetic energy spectra for the non--mesonic
%weak decay of $^{12}_\Lambda$C. The total spectrum $N_p$ (normalized
%{\it per non--mesonic weak decay}) has been decomposed
%in its components $N^{\Lambda n\to nn}_p$,
%$N^{\Lambda p\to np}_p$ and $N^{\Lambda np\to nnp}_p$ according to
%Eq.~(\ref{single-n}).}
\label{s4p}
\end{center}
\end{figure}

\begin{figure}
\begin{center}
\mbox{\epsfig{file=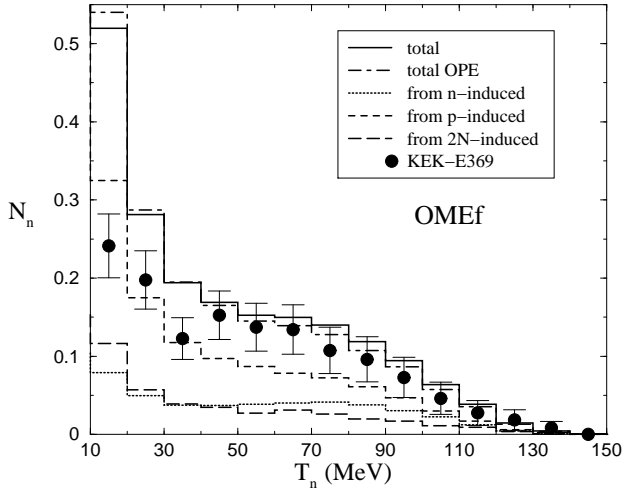,width=.45\textwidth}}
%\mbox{\epsfig{file=s4n.eps}}
\vskip 2mm
\caption{Same as in Fig.~\ref{s3n} for $^{12}_\Lambda$C.
The OPE result is also shown.
Experimental data are taken from Ref.~\protect\cite{Ki02}.}
\label{s4n}
\end{center}
\end{figure}

In Figs.~\ref{s3p} and \ref{s3n}, $N_N$, $N^{\Lambda n\to nn}_N$, $N^{\Lambda p\to np}_N$ and
$N^{\Lambda np\to nnp}_N$ are shown as functions of the nucleon kinetic energy
in the case of $^5_\Lambda$He.
In Figs.~\ref{s4p} and \ref{s4n} the same quantities are given for 
$^{12}_\Lambda$C. From the
$N_n$ and $N_p$ distributions for $^5_\Lambda$He we note that the 
maxima occurring at $T_N\simeq 75$ MeV  ---mainly due to 
the kinematics of the weak decay nucleons (see the dashed line in
Fig.~\ref{s1} for protons)---
are more pronounced for neutrons than for protons. Note that for any value of the
ratio $\Gamma_n/\Gamma_p$,
the number of primary neutrons is indeed larger than the number of 
primary protons [see Eq.~(\ref{ratio-np1})]. Consequently, due to neutron--proton reactions,
the proportion of secondary protons in the proton spectrum is larger
than the proportion of secondary neutrons in the neutron spectrum.
From Figs.~\ref{s3p} and \ref{s4p} we also note that the fractions of 
\emph{protons} emitted in {\it neutron}--induced processes are quite small.
For values of $T_p$ in the $30\div 40$ MeV bin (i.e., just above the experimental 
threshold), $N^{\Lambda n\to nn}_p/N^{\Lambda p\to np}_p\simeq 0.15$ both for 
$^5_\Lambda$He and $^{12}_\Lambda$C, which corresponds to an interference term 
$I_p\, {\rm cos}\, \phi_p=0.67\, {\rm cos}\, \phi_p$ in 
Eq.~(\ref{single}). Therefore, even if the number of protons from $n$--induced decays
is expected to be rather small, the existence of such events could nevertheless 
produce a quite big interference effect: unfortunately, the matter depends 
crucially on the phase $\phi_p$ which we cannot estimate theoretically. 
From Figs.~\ref{s3p} and \ref{s4p} one also note that the two--nucleon induced
mechanism could be responsible for an
interference effect in the proton energy spectra as important as the one
which could be due to the $n$--induced channel. For single--{\it neutron} spectra
these interference effects could be even larger. Indeed,
from Figs.~\ref{s3p}--\ref{s4n} it is clear that, as expected, 
$N^{\Lambda n\to nn}_p/N^{\Lambda p\to np}_p<
N^{\Lambda n\to nn}_n/N^{\Lambda p\to np}_n<1$.
Once again this conclusion neglects the effect of the phases of the
interfering amplitudes. 


The single--neutron spectrum for $^{12}_\Lambda$C observed in the KEK--E369 experiment 
\cite{Ki02} is well reproduced by our calculations. This is evident from 
Fig.~\ref{s4n}, where we show results based on two models (OPE and OMEf)
which predict quite different $\Gamma_n/\Gamma_p$ ratios. Unfortunately,
the dependence of the neutron spectra on variations of $\Gamma_n/\Gamma_p$ is very weak
(the same is true also for the proton spectra)
and a precise extraction of the ratio from the KEK--E369 distribution is not possible. 
We checked that an analogous calculation performed with the polarization propagator
method in local density approximation of Ref.~\cite{Ra02} supplies neutron spectra which 
reproduce the data of Fig.~\ref{s4n} with a $\chi^2$ per D.O.F.~ smaller than 1
when $\Gamma_n/\Gamma_p$ (a free parameter in such kind
of calculations) is chosen to lie in the range $0\div1.5$ and 
data above $T_n=30$ MeV are considered.
The little sensitivity of $N_n$ and $N_p$ to $\Gamma_n/\Gamma_p$ is mainly due to
the fact that these numbers are normalized per non--mesonic weak decay.   
As a consequence, one should separately compare the complementary observable,
$\Gamma_n+\Gamma_p$, with the experiment. For $^{12}_\Lambda$C, our calculations supply 
$\Gamma_{\rm NM}\equiv \Gamma_n+\Gamma_p+\Gamma_2=
1.25\,(\Gamma_n+\Gamma_p)=0.91$ or $0.69$ when model OMEa or OMEf is
used. These values agree quite well with the KEK data $0.83\pm 0.11$ of Ref.~\cite{Ou00}.

The problem of the small sensitivity of $N_n$ and $N_p$
to variations of $\Gamma_n/\Gamma_p$ can be overcome if one concentrates 
on another single--nucleon observable.
The ratio $\Gamma_n/\Gamma_p$ is defined in terms of the ratio
between the number of primary weak decay neutrons and protons,
$N^{\rm wd}_n$ and $N^{\rm wd}_p$, in the following way:
\begin{equation}
\label{ratio-np1}
\frac{\Gamma_n}{\Gamma_p}\equiv \frac{1}{2}\left(\frac{N^{\rm wd}_n}{N^{\rm wd}_p}-1\right) .
\end{equation} 
Due to two--body induced decays and (especially) nucleon FSI, one expects the inequality:
\begin{equation}
\label{ratio-np2}
\frac{\Gamma_n}{\Gamma_p}
\neq \frac{1}{2}\left(\frac{N_n}{N_p}-1\right)\equiv
R_1\left[\Delta T_n, \Delta T_p, \Gamma_2\right]
\end{equation}
to be valid in a situation, such as the experimental one, in which particular intervals
of variability of the neutron and proton kinetic energy,
$\Delta T_n$ and $\Delta T_p$, are employed in the determination of $N_n$ and $N_p$.

As one can deduce from previous figures,
not only $N_n$ and $N_p$ but also the ratio $N_n/N_p$ depends
on $\Delta T_n$ and $\Delta T_p$. This is more evident from Table \ref{erre}, in which
the function $R_1$ defined by Eq.~(\ref{ratio-np2}) is given
for $^5_\Lambda$He and $^{12}_\Lambda$C, for
different nucleon energy thresholds $T^{\rm th}_N$ and for the OPE, OMEa and OMEf models. 
For a given energy threshold, $R_1$ is closer to $\Gamma_n/\Gamma_p$ 
for $^5_\Lambda$He than for $^{12}_\Lambda$C since FSI are larger
in carbon. The ratio $N_n/N_p$ (or equivalently $R_1$)
is more sensitive to variations of $\Gamma_n/\Gamma_p$
(see the differences between the OPE, OMEa and OMEf calculations of Table \ref{erre})
than $N_n$ and $N_p$ separately. Moreover, $N_n/N_p$ is less affected by 
FSI than $N_n$ and $N_p$. Therefore, measurements of
$N_n/N_p$ should permit to determine $\Gamma_n/\Gamma_p$
with better precision. The recent KEK--E462 experiment has measured
the ratio $N_n/N_p$ for $^5_\Lambda$He: a preliminary analysis of the data supplies
a value of $R_1$ around 0.6 with a relative error of about 20\% 
using nucleon energy thresholds of $50$ and $60$ MeV \cite{Bh03}. 
Our result of Table~\ref{erre} corresponding to the OMEf calculation for $T^{\rm th}_N=60$ 
MeV agree with this experimental determination. According to this comparison, 
the ratio $\Gamma_n/\Gamma_p$ for $^5_\Lambda$He should be around the value 
(0.46) predicted by the OMEf model.

%*******************************************
\subsection{Double--coincidence spectra}
%*******************************************
\label{res-coin}

\begin{figure}
\begin{center}
\mbox{\epsfig{file=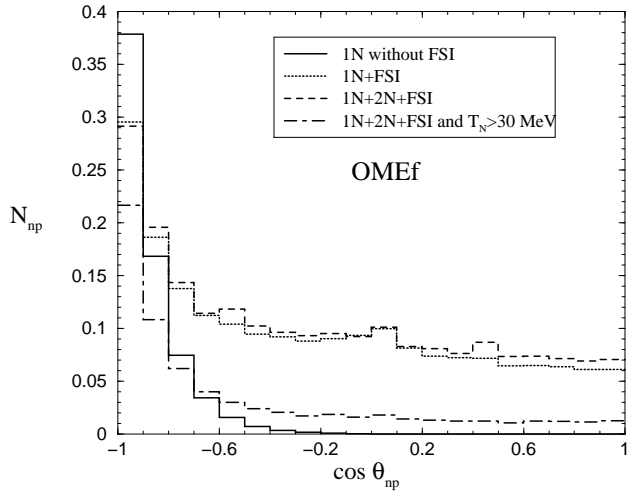,width=.45\textwidth}}
%\mbox{\epsfig{file=0.eps}}
\vskip 2mm
\caption{Opening angle distribution of $np$ pairs
emitted in the non--mesonic decay of $^5_\Lambda$He.
The continuous and dotted lines are normalized 
{\it per one--nucleon induced decay} ($\Gamma_1=\Gamma_n+\Gamma_p$), while
the dashed and dot--dashed curves are normalized {\it per non--mesonic 
decay} ($\Gamma_{\rm NM}=\Gamma_1+\Gamma_2$).}
\label{0}
\end{center}
\end{figure}

In Fig.~\ref{0} we report the opening angle distribution of $np$ pairs
emitted in the non--mesonic decay of $^5_\Lambda$He. 
Note the particular normalizations of the curves presented in this and in the
following figures.
Neglecting nucleon FSI and the two--nucleon induced channel, the angular correlation
is strongly peaked at $\theta_{np}=180^\circ$ and only 1 pair
out of 100 is emitted with an opening angle smaller than $115^\circ$
(i.e., with cos $\theta_{np} \geq -0.4$). 
The effect of FSI is to decrease the weak decay nucleon back--to--back 
distribution (i.e., for cos $\theta_{np}\leq -0.9$) of about 25\%
and populate, strongly, the spectrum for cos $\theta_{np}\geq -0.8$.
The effect of the two--nucleon induced channel is moderate: practically, 
it only increases the distribution (by about 20\%) in the region 
with cos $\theta_{np}>0.4$. When a nucleon kinetic
energy cut of $30$ MeV is applied, large part of the distribution
is removed. This is due to FSI, which leads to a large amount of pairs 
($\simeq 70$\% of the total) in which at least one nucleon
has kinetic energy below $30$ MeV. 

\begin{figure}
\begin{center}
\mbox{\epsfig{file=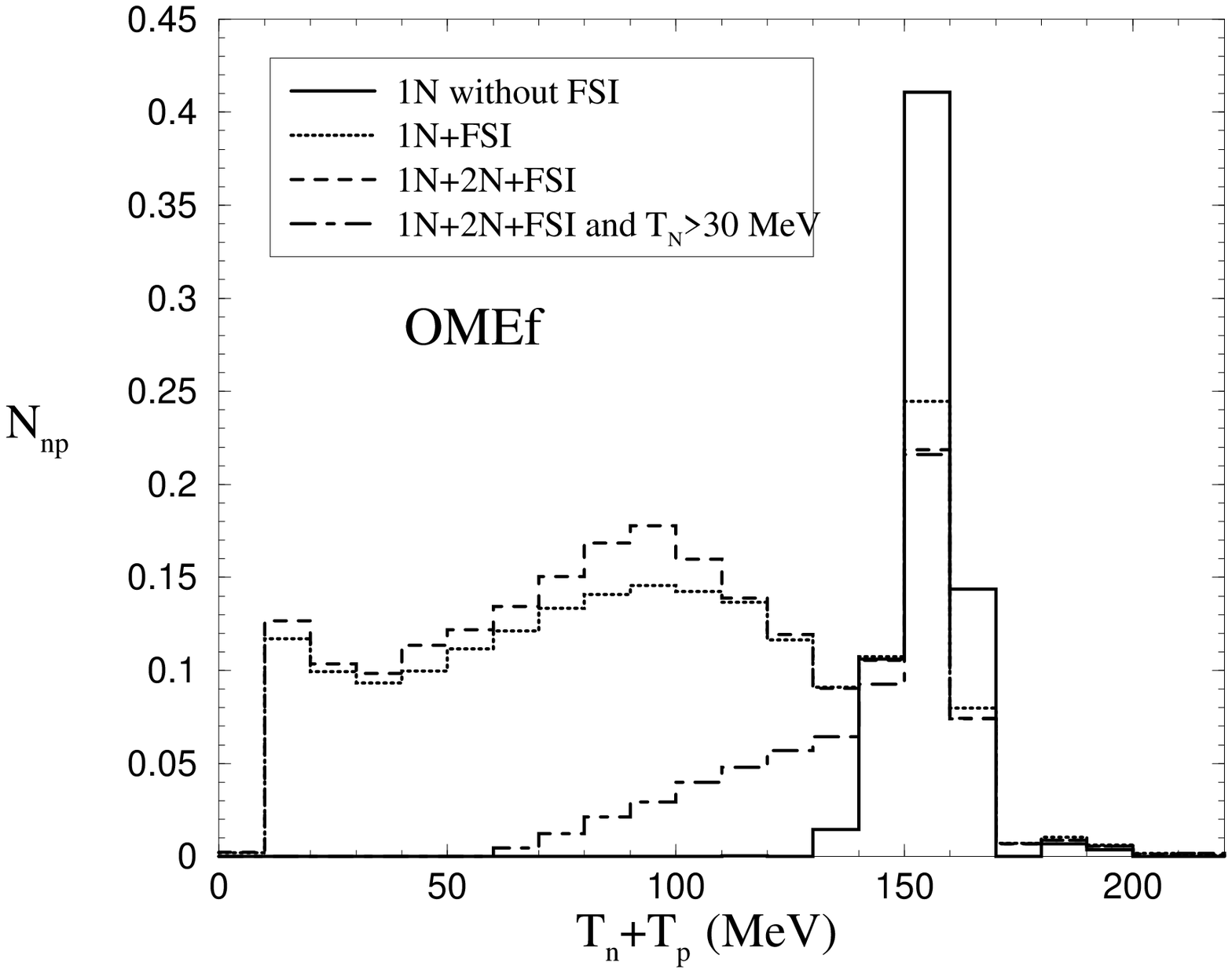,width=.45\textwidth}}
%\mbox{\epsfig{file=01.eps}}
\vskip 2mm
\caption{Kinetic energy sum distribution of $np$ pairs
emitted in the non--mesonic decay of $^5_\Lambda$He.
The continuous and dotted lines are normalized
{\it per one--nucleon induced decay} ($\Gamma_1=\Gamma_n+\Gamma_p$), while
the dashed and dot--dashed curves are normalized {\it per non--mesonic
decay} ($\Gamma_{\rm NM}=\Gamma_1+\Gamma_2$).}
\label{01}
\end{center}
\end{figure}

In Fig.~\ref{01} we show the kinetic energy correlation of $np$ pairs
emitted in the decay of $^5_\Lambda$He. 
Neglecting nucleon FSI and the two--nucleon induced channel, the energy correlation
is strongly peaked, as expected, at $T_n+T_p\simeq 155$ MeV. Indeed,
the $Q$--value corresponding to the proton--induced three--body process
$^5_\Lambda{\rm He}\to$~$^3{\rm H}+n+p$ is 153 MeV. 
The effect of the FSI is to decrease the back--to--back
maximum and to populate, strongly, the spectrum for 
$T_n+T_p \leq 140$ MeV. The effect of the two--nucleon induced channel is only visible
when $T_n+T_p$ is below $110$ MeV, where it enhances the distribution. 
Once a $30$ MeV kinetic energy cut is applied, the distribution at small
$T_n+T_p$ is considerably reduced for the same reason explained in the
previous paragraph.

The opening angle and kinetic energy correlations for $nn$ pairs have essentially 
the same structure of the $np$ distributions showed in Figs.~\ref{0} and \ref{01}.
For a discussion of the $pp$ distributions and the different effect of FSI 
in $^5_\Lambda$He and $^{12}_\Lambda$C we refer to our previous paper \cite{previous}.

The ratio $\Gamma_n/\Gamma_p$ is defined as the ratio
between the number of weak decay $nn$ and $np$ pairs,
$N^{\rm wd}_{nn}$ and $N^{\rm wd}_{np}$. However, due to 
two--body induced decays and (especially) nucleon FSI effects, one has:
\begin{equation}
\label{ratio-nn}
\frac{\Gamma_n}{\Gamma_p}\equiv \frac{N^{\rm wd}_{nn}}{N^{\rm wd}_{np}}
\neq \frac{N_{nn}}{N_{np}}=R_2\left[\Delta \theta_{12}, \Delta T_n, \Delta T_p, \Gamma_2\right] 
\end{equation}
when $N_{nn}$ and $N_{np}$ are determined by employing particular intervals 
of variability of the pair opening angle, $\Delta \theta_{12}$, and the nucleon
kinetic energies, $\Delta T_n$ and $\Delta T_p$.
Actually, as one can deduce from the figures we discussed in Ref.~\cite{previous},
not only $N_{nn}$ and $N_{np}$ but also the ratio $N_{nn}/N_{np}$ depends
on $\Delta \theta_{12}$, $\Delta T_n$ and $\Delta T_p$. The discussion
of Ref.~\cite{previous} also proves that the ratio $N_{nn}/N_{np}$ is much less sensitive
to FSI effects and variations of the energy
cuts and angular restrictions than $N_{nn}$ and $N_{np}$ separately.    

The numbers of nucleon pairs $N_{NN}$ discussed up to this point
are related to the corresponding quantities for the one--nucleon ($N^{\rm 1B}_{NN}$)
and two--nucleon ($N^{\rm 2B}_{NN}$) induced processes
[the former (latter) being normalized per one--body (two--body) stimulated
non--mesonic weak decay] via the following equation:
\begin{eqnarray}
\label{1-2}
N_{NN}&=&\frac{N^{\rm 1B}_{NN}\, \Gamma_1+N^{\rm 2B}_{NN}\, \Gamma_2}{\Gamma_1+\Gamma_2}  \\
&\equiv& N^{\Lambda n\to nn}_{NN}+
N^{\Lambda p\to np}_{NN}+N^{\Lambda np\to nnp}_{NN}, \nonumber
\end{eqnarray} 
where:
\begin{eqnarray}
\label{1-22}
N^{\rm 1B}_{NN}&=&\frac{N^{\rm 1Bn}_{NN}\, \Gamma_n+N^{\rm 1Bp}_{NN}\, 
\Gamma_p} {\Gamma_1} ,
\end{eqnarray}  
and the remaining $N$'s have obvious meaning. Therefore, the quantities
$N^{\rm 1Bn}_{NN}$, $N^{\rm 1Bp}_{NN}$ and $N^{\rm 2B}_{NN}$
($N^{\Lambda n\to nn}_{NN}$, $N^{\Lambda p\to np}_{NN}$ and $N^{\Lambda np\to nnp}_{NN}$) 
{\it do not} depend ({\it do} depend) on the interaction model
employed to describe the weak decay. 

In Table~\ref{np-contr} we report the OPE, OMEa and OMEf predictions for $N_{np}$
and its components $N^{\Lambda n\to nn}_{np}$, $N^{\Lambda p\to np}_{np}$ 
and $N^{\Lambda np\to nnp}_{np}$
in the case of $^{12}_\Lambda$C. Two different
opening angle regions and an energy threshold of $30$ MeV have been 
considered. The contribution to $N_{np}$ of the $n$--induced decay channel is always 
smaller than or comparable with the one coming from the two--nucleon induced mechanism: 
$N^{\Lambda n\to nn}_{np}\lsim N^{\Lambda np\to nnp}_{np}$. The ratio 
$N^{\Lambda n\to nn}_{np}/N^{\Lambda p\to np}_{np}$ is always smaller than $0.10$,
which corresponds to $I_{np}< 0.57$ in Eq.~(\ref{double}). This shows that interferences 
in coincidence observables are potentially smaller than in single--nucleon 
spectra but also that they might be 
non--negligible. Due to the less pronounced
effects of FSI in $^5_\Lambda$He, smaller values of $I_{np}$ have been obtained 
in this second case.

\begin{figure}
\begin{center}
\mbox{\epsfig{file=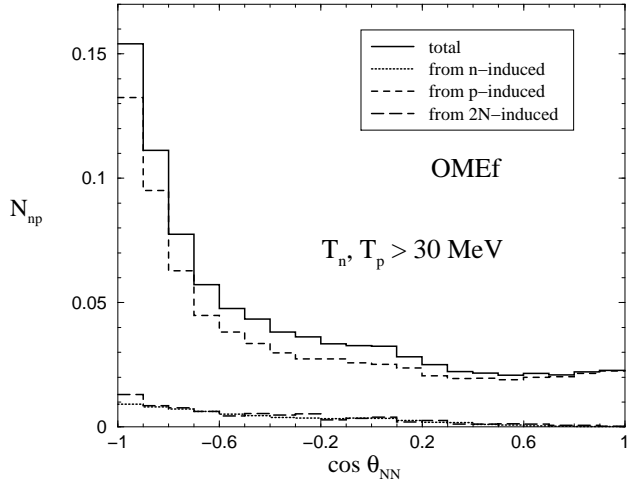,width=.45\textwidth}}
%\mbox{\epsfig{file=02.eps}}
\vskip 2mm
\caption{Opening angle distributions of $np$ pairs emitted
in the non--mesonic decay of $^{12}_\Lambda$C.
The total spectrum $N_{np}$ (normalized
{\it per non--mesonic weak decay}) has been decomposed
in its components $N^{\Lambda n\to nn}_{np}$,
$N^{\Lambda p\to np}_{np}$ and $N^{\Lambda np\to nnp}_{np}$ according to
Eqs.~(\ref{1-2}) and (\ref{1-22}).}
\label{02}
\end{center}
\end{figure}

\begin{figure}
\begin{center}
\mbox{\epsfig{file=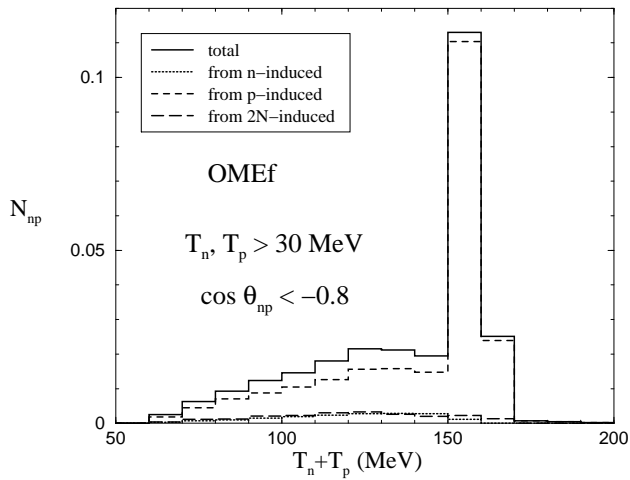,width=.45\textwidth}}
%\mbox{\epsfig{file=03.eps}}
\vskip 2mm
\caption{Kinetic energy correlations of $np$ pairs emitted
in the non--mesonic decay of $^{12}_\Lambda$C at cos $\theta_{np}\leq -0.8$.
The total spectrum $N_{np}$ (normalized
{\it per non--mesonic weak decay}) has been decomposed
in its components $N^{\Lambda n\to nn}_{np}$,
$N^{\Lambda p\to np}_{np}$ and $N^{\Lambda np\to nnp}_{np}$ according to
Eqs.~(\ref{1-2}) and (\ref{1-22}).}
\label{03}
\end{center}
\end{figure}

In Fig.~\ref{02} we show the $np$ pair opening angle distribution in the 
case of $^{12}_\Lambda$C. The total spectrum $N_{np}$ has been decomposed
into the components $N^{\Lambda n\to nn}_{np}$, $N^{\Lambda p\to np}_{np}$ 
and $N^{\Lambda np\to nnp}_{np}$.
A nucleon energy threshold of $30$ MeV has been used in the calculation.
Fig.~\ref{03} corresponds to the kinetic energy correlation of $np$ pairs:
it is again for $^{12}_\Lambda$C and $T^{\rm th}_N=30$ MeV, but now only
back--to--back angles (${\rm cos}\, \theta_{np}\leq -0.8$) have been taken into account. 
We note how both the $n$--induced and the two--nucleon induced decay 
processes give very small contributions to the total distributions in Figs.~\ref{02} and \ref{03}. 
Nevertheless, these decay processes could produce non--negligible interference terms.
To minimize this effect, one could consider, for instance, 
not only back--to--back angles but also nucleon kinetic energies 
in the interval $150\div170$ MeV, for which we predict $I_{np}=0.18$.
Again, smaller values of $I_{np}$ have been obtained for $^5_\Lambda$He.


In Table \ref{ome-che} the ratio $N_{nn}/N_{np}$ predicted by the
OPE, OMEa and OMEf models for $^5_\Lambda$He 
and $^{12}_\Lambda$C is given for two opening angle intervals
and for a nucleon energy threshold of $30$ MeV. The results of the OMEa and OMEf
models are in reasonable agreement with the preliminary KEK--E462 data 
for $^5_\Lambda$He \cite{OuOk}.

This data ($N_{nn}/N_{np}=0.44\pm0.11$), which
corresponds to an energy threshold of $30$ MeV
and cos $\theta_{np}\leq -0.8$, can be fitted using the 6
weak interaction model independent quantities
$N^{\rm 1Bn}_{nn}$, $N^{\rm 1Bp}_{nn}$, $N^{\rm 2B}_{nn}$,
$N^{\rm 1Bn}_{np}$, $N^{\rm 1Bp}_{np}$ and $N^{\rm 2B}_{np}$ entering 
Eqs.~(\ref{1-2}) and (\ref{1-22}) and quoted in Table~\ref{univ}. 
This can be achieved through the following relation:
\begin{equation}
\label{fit}
\frac{N_{nn}}{N_{np}}=\frac{\displaystyle\left(N^{\rm 1Bn}_{nn}+N^{\rm 2B}_{nn}
\frac{\Gamma_2}{\Gamma_1}\right) \frac{\Gamma_n}{\Gamma_p}
+N^{\rm 1Bp}_{nn}+N^{\rm 2B}_{nn}\frac{\Gamma_2}{\Gamma_1}}
{\displaystyle\left(N^{\rm 1Bn}_{np}+N^{\rm 2B}_{np} \frac{\Gamma_2}{\Gamma_1}\right)
\frac{\Gamma_n}{\Gamma_p}
+N^{\rm 1Bp}_{np}+N^{\rm 2B}_{np}\frac{\Gamma_2}{\Gamma_1}} , 
\end{equation}
using $\Gamma_n/\Gamma_p$ and $\Gamma_2/\Gamma_1$ as fitting parameters.

\begin{figure}
\begin{center}
\mbox{\epsfig{file=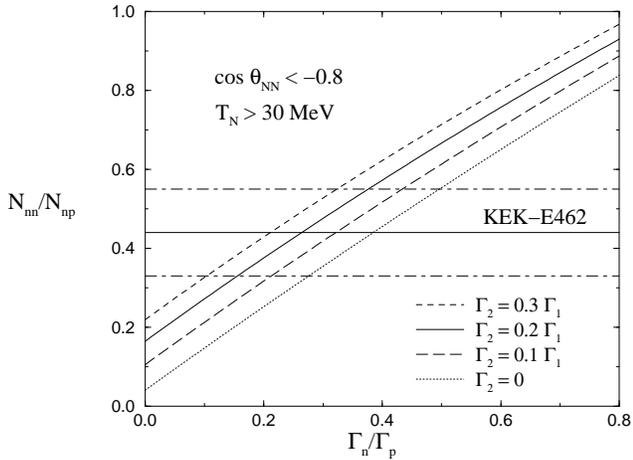,width=.45\textwidth}}
%\mbox{\epsfig{file=fithe.eps}}
\vskip 2mm
\caption{Dependence of the ratio $N_{nn}/N_{np}$ on $\Gamma_n/\Gamma_p$
and $\Gamma_2/\Gamma_1$ for $^5_\Lambda$He.
The results correspond to a nucleon energy threshold of $30$ MeV and
the angular restriction cos~$\theta_{NN}\leq -0.8$.
The horizontal lines show the
preliminary KEK--E462 data of Ref.~\protect\cite{OuOk}.}
\label{fithe}
\end{center}
\end{figure}

\begin{figure}
\begin{center}
\mbox{\epsfig{file=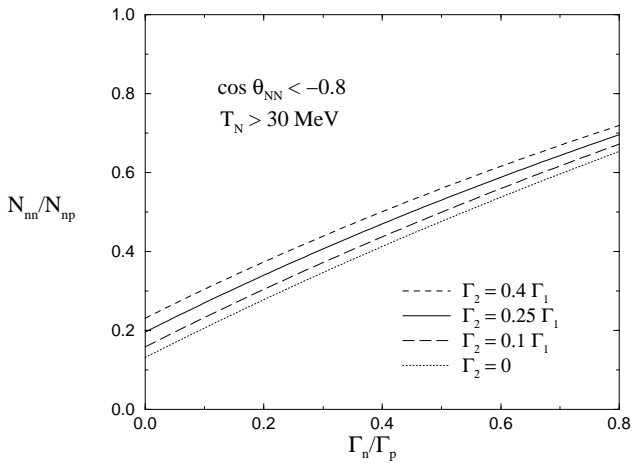,width=.45\textwidth}}
%\mbox{\epsfig{file=fitc.eps}}
\vskip 2mm
\caption{Same as in Fig.~\ref{fithe} for $^{12}_\Lambda$C.}
%\caption{Dependence of the ratio $N_{nn}/N_{np}$ on $\Gamma_n/\Gamma_p$
%and $\Gamma_2/\Gamma_1$ for $^{12}_\Lambda$C.
%The results correspond to a nucleon energy threshold of $30$ MeV and
%athe angular restriction cos~$\theta_{NN}\leq -0.8$.}
\label{fitc}
\end{center}
\end{figure}

In Fig.~\ref{fithe} (Fig.~\ref{fitc}) we report the dependence of $N_{nn}/N_{np}$ for
$^5_\Lambda$He ($^{12}_\Lambda$C) on the ratio
$\Gamma_n/\Gamma_p$ for four different values of $\Gamma_2/\Gamma_1$.
Both figures correspond
to the case with a nucleon energy threshold of $30$ MeV and the angular restriction
cos $\theta_{np}\leq -0.8$. For a given value of $\Gamma_2/\Gamma_1$,
Figs.~\ref{fithe} and \ref{fitc} permit
an immediate determination of $\Gamma_n/\Gamma_p$ by a direct comparison with
data for the observable $N_{nn}/N_{np}$.  

Using the $^5_\Lambda {\rm He}$ data $N_{nn}/N_{np}=0.44\pm0.11$ and
assuming $\Gamma_2=0$, from Eq.~(\ref{fit}) we obtain the following fitted ratio:
\begin{equation}
\label{fit1}
\frac{\Gamma_n}{\Gamma_p}\left(^5_\Lambda {\rm He}\right)=0.39\pm0.11 .
\end{equation}
By employing the value $\Gamma_2/\Gamma_1=0.2$ (i.e., the one obtained with the model
of Ref.~\cite{Al00} and used in our calculations), a 34\% reduction of the ratio
is predicted:
\begin{equation}
\label{fit2}
\frac{\Gamma_n}{\Gamma_p}\left(^5_\Lambda {\rm He}\right)=0.26\pm 0.11 .
\end{equation} 
These values of $\Gamma_n/\Gamma_p$ are rather small if compared
with previous determinations \cite{Sz91} ($0.93\pm 0.55$)
\cite{No95a} ($1.97\pm 0.67$) from single--nucleon energy spectra analyses. 
On the contrary, the results of Eqs.~(\ref{fit1}) and (\ref{fit2}) are
in agreement with the pure theoretical predictions of Refs.~\cite{Pa01,It98,Ok99}.

Forthcoming coincidence data from KEK and FINUDA
could be directly compared with the results discussed in this paper
and will permit to achieve new determinations of the $\Gamma_n/\Gamma_p$ ratio
and to establish the first constraints on $\Gamma_2/\Gamma_1$.     

%**************************
\section{Conclusions}
%**************************
\label{conc}

In this work we have presented a calculation of
single and double--coincidence observables for the nucleons 
emitted in the non--mesonic weak decay of $\Lambda$--hypernuclei.
This has been possible by supplementing our OME weak interaction models
with FSI through an intranuclear cascade calculation.

The results of our calculations
are in reasonable agreement with preliminary KEK--E462 data for the 
observable ratio $N_{nn}/N_{np}$ in $^5_\Lambda$He. 

We also perform a weak interaction model independent analysis in which  
$\Gamma_n/\Gamma_p$ and $\Gamma_2/\Gamma_1$ are considered as free parameters: 
the KEK $^5_\Lambda {\rm He}$ data $N_{nn}/N_{np}=0.44\pm 0.11$ is reproduced if
$\Gamma_n/\Gamma_p=0.39\pm 0.11$ and $\Gamma_2=0$ 
or $\Gamma_n/\Gamma_p=0.26\pm 0.11$ and $\Gamma_2/\Gamma_1=0.2$.
%Forthcoming coincidence data from KEK and FINUDA
%could be directly compared with the results discussed in this paper
%and will possibly permit to determine the $\Gamma_n/\Gamma_p$ ratio with better precision
%and to establish the first constraints on $\Gamma_2/\Gamma_1$. 

The extension of the
present study to triple--nucleon coincidence is of interest both for 
the determination of $\Gamma_n/\Gamma_p$ and to disentangle
the effects of one-- and two--nucleon induced decay channels.

The values of $\Gamma_n/\Gamma_p$ we obtain by fitting 
KEK coincidence data for $^5_\Lambda$He are in agreement
with other recent theoretical evaluations. However, they are
rather small if compared with the results of previous analyses
from single--nucleon energy spectra. 
Actually, all the previous experimental analyses of single--nucleon
spectra \cite{Mo74,Sz91,Ha01,Sato02,No95a,No95}, supplemented in some cases by
intranuclear cascade calculations, derived $\Gamma_n/\Gamma_p$ values in
disagreement with pure theoretical predictions. In our opinion,
the fact that our calculations reproduce coincidence data
for values of $\Gamma_n/\Gamma_p$ as small as $0.3\div 0.4$
could signal the existence of non--negligible interference effects between
the $n$-- and $p$--induced channels in those old single--nucleon data.

In conclusion, although further (theoretical and experimental) confirmation is 
needed, we think that our investigation proves how the study of nucleon
coincidence observables can offer a promising possibility
to solve the longstanding puzzle on the $\Gamma_n/\Gamma_p$ ratio.

\vskip 4mm
This work is partly supported by EURIDICE HPRN--CT--2002--00311,
by the DGICYT BFM2002--01868, by the Generalitat de Catalunya SGR2001--64 and by
INFN. Discussions with O. Hashimoto, H. Outa and Y. Sato are acknowledged.

%\vskip 4cm

\begin{table}
\begin{center}
\caption{Weak decay rates (in units of the free $\Lambda$ decay width) 
predicted by Ref.~\protect\cite{Pa01}
for $^5_\Lambda$He and $^{12}_\Lambda$C.}
\label{gammas}
\begin{tabular}{c|c c c|c c c}
\mc {1}{c|}{} &
\mc {1}{c}{} &
\mc {1}{c}{$\Gamma_n+\Gamma_p$} &
\mc {1}{c|}{} &
\mc {1}{c}{} &
\mc {1}{c}{$\Gamma_n/\Gamma_p$} &   
\mc {1}{c}{} \\
                  &   OPE  &  OMEa  &  OMEf &   OPE  & OMEa   & OMEf   \\ \hline
$^5_\Lambda$He    & $0.43$ & $0.43$ & $0.32$ & $0.09$ & $0.34$ & $0.46$   \\
$^{12}_\Lambda$C  & $0.75$ & $0.73$ & $0.55$ & $0.08$ & $0.29$ & $0.34$   \\ 
\end{tabular}
\end{center}
\end{table}
    
\vskip -7mm
\begin{table}
\begin{center}
\caption{Predictions for the quantity $R_1$ of Eq.~(\ref{ratio-np2}) for 
$^5_\Lambda$He and $^{12}_\Lambda$C corresponding to different nucleon thresholds $T^{\rm th}_N$
and to the OPE, OMEa and OMEf models.}
\label{erre}
\begin{tabular}{c c|c c c|c}
\mc {1}{c}{} &  
\mc {1}{c|}{} &
\mc {1}{c}{} &
\mc {1}{c}{$T^{\rm th}_N$ (MeV)} & 
\mc {1}{c|}{} &
\mc {1}{c}{} \\
                 &          & $0$     & $30$    & $60$   & $\Gamma_n/\Gamma_p$ \\ \hline
                 & OPE      & $0.04$  & $0.13$  & $0.16$ & $0.09$  \\ 
$^5_\Lambda$He   & OMEa     & $0.15$  & $0.32$  & $0.39$ & $0.34$  \\ 
                 & OMEf   & $0.19$  & $0.40$  & $0.49$ & $0.46$ \\ \hline
                 & OPE      & $-0.06$ & $-0.01$ & $0.05$ & $0.08$ \\   
$^{12}_\Lambda$C & OMEa   & $-0.02$ & $0.07$  & $0.19$ & $0.29$ \\
                 & OMEf   & $-0.01$ & $0.09$  & $0.21$ & $0.34$ \\  
\end{tabular} 
\end{center} 
\end{table}

\vskip -7mm 
\begin{table}
\begin{center}
\caption{OPE, OMEa and OMEf predictions for $N_{np}$ and its components
$N^{\Lambda n\to nn}_{np}$, $N^{\Lambda p\to np}_{np}$ and
$N^{\Lambda np\to nnp}_{np}$ (integrated over all angles and for energies
$T_N\geq 30$ MeV) as given by Eqs.~(\ref{1-2}) and (\ref{1-22})
for the case of $^{12}_\Lambda$C . The numbers in parentheses
correspond to the angular region with cos~$\theta_{np}\leq -0.8$.}
\label{np-contr}
\begin{tabular}{c|c c c c}
\mc {1}{c|}{} &
\mc {1}{c}{$N_{np}$} &
\mc {1}{c}{$N^{\Lambda n\to nn}_{np}$} &
\mc {1}{c}{$N^{\Lambda p\to np}_{np}$} &
\mc {1}{c}{$N^{\Lambda np\to nnp}_{np}$} \\ \hline
OPE     & $1.00$ ($0.32$) & $0.02$ ($0.00$) & $0.91$ ($0.29$)  & $0.08$ ($0.02$) \\ 
OMEa    & $0.89$ ($0.27$) & $0.06$ ($0.02$) & $0.76$ ($0.24$)  & $0.08$ ($0.02$) \\
OMEf    & $0.87$ ($0.27$) & $0.07$ ($0.02$) & $0.73$ ($0.23$)  & $0.08$ ($0.02$) \\
\end{tabular}
\end{center}
\end{table}     

\vskip -7mm 
\begin{table}
\begin{center}
\caption{Predictions for the ratio $R_2\equiv N_{nn}/N_{np}$
for $^5_\Lambda$He and $^{12}_\Lambda$C.
An energy thresholds $T^{\rm th}_N$ of $30$ MeV and two pair opening angle
regions have been considered. The (preliminary) data are from KEK--E462 \protect\cite{OuOk}.}
\label{ome-che}
\begin{tabular}{c|c c | c c}
\mc {1}{c|}{} &
\mc {1}{c}{$^5_\Lambda$He} &
\mc {1}{c|}{} &
\mc {1}{c}{$^{12}_\Lambda$C} &
\mc {1}{c}{} \\
       & cos $\theta_{NN}\leq -0.8$ & all $\theta_{NN}$ & 
cos $\theta_{NN}\leq -0.8$ & all $\theta_{NN}$ \\ \hline
OPE     & $0.25$  & $0.26$  & $0.24$ & $0.29$  \\ 
OMEa    & $0.51$  & $0.45$  & $0.39$ & $0.37$  \\
OMEf    & $0.61$  & $0.54$  & $0.43$ & $0.39$ \\ \hline
EXP  & $0.44\pm 0.11$  &     &        &  \\   
\end{tabular}
\end{center}
\end{table} 

\vskip -7mm 
\begin{table}[b]
\begin{center}
\caption{Predictions for the weak interaction model independent quantities
$N^{\rm 1Bn}_{NN}$, $N^{\rm 1Bp}_{NN}$ and
$N^{\rm 2B}_{NN}$ (integrated over all angles and for energies
$T_N\geq 30$ MeV) of Eqs.~(\ref{1-2}) and (\ref{1-22})
for $^5_\Lambda$He and $^{12}_\Lambda$C. The numbers in parentheses
correspond to the angular region with cos~$\theta_{NN}\leq -0.8$.}
\label{univ}
\begin{tabular}{c|c c c}
\mc {1}{c|}{} &
\mc {1}{c}{$N^{\rm 1Bn}_{nn}$} &
\mc {1}{c}{$N^{\rm 1Bp}_{nn}$} &
\mc {1}{c}{$N^{\rm 2B}_{nn}$} \\ \hline
$^5_\Lambda$He    & $0.84$ ($0.53$) & $0.10$ ($0.02$) & $0.54$ ($0.34$)   \\
$^{12}_\Lambda$C  & $0.56$ ($0.30$) & $0.27$ ($0.05$) & $0.30$ ($0.12$)   \\ \hline\hline
   & $N^{\rm 1Bn}_{np}$ & $N^{\rm 1Bp}_{np}$ & $N^{\rm 2B}_{np}$ \\ \hline
$^5_\Lambda$He    & $0.20$ ($0.05$) & $0.98$ ($0.49$) & $0.55$ ($0.22$)   \\
$^{12}_\Lambda$C  & $0.33$ ($0.08$) & $1.22$ ($0.38$) & $0.38$ ($0.11$)   \\
\end{tabular}
\end{center}
\end{table}   

\end{multicols}
\end{document}